\documentclass[doublecol]{epl2}

\usepackage{graphicx}
\usepackage{amssymb}
\newcommand{\beq}{\begin{equation}}
\newcommand{\enq}{\end{equation}}
\newcommand{\bea}{\begin{eqnarray}}
\newcommand{\ena}{\end{eqnarray}}

\begin{document}

\title{Towards a Bose glass transition in an optical Penrose
  quasicrystal} \author{Alberto Cetoli and Emil Lundh}
\institute{Department of Physics, Ume{\aa} University, \ SE-90187
  Ume{\aa}, Sweden} \abstract{We study numerically a 2D Bose-Einstein 
  condensate in a quasiperiodic array of potential peaks, assumed 
  to be generated by superimposing five blue detuned laser beams.
  By using a Bogoliubov \emph{ansatz} for the excitations we show that
  the system approaches a gapless, insulating phase upon increasing
  the potential, consistent with a Bose glass phase. 
  The characteristics of the transition in terms of phase correlations, 
  oscillatory modes, and superfluid fraction are discussed.
}

\maketitle

\section{Introduction}
The existence of a quasiperiodic structure was first observed in
a rapidly quenched metal alloy by Shechtman {\it et al.} 
\cite{Shechtman}. A sharp peak in the Bragg scattering proved
the existence of long range order. The system, however, lacked
translational symmetry, undermining the established belief that an
ordered structure must be periodic. This novel configuration was 
generalized by Levine and Steinhardt \cite{levine}, who formalized
the concept of quasiperiodic crystals (QC): these formations exhibit 
long range order without having translational symmetry, acting both
as crystals, admitting a metal-insulator transition, and as amorphous
solids, with peculiar localization properties and fractal spectrum
\cite{QPbook}.

In cold atom research, the investigation of the effects of 
repulsion between bosons when a disordered potential is present is an
ongoing quest. In the late 80s it was speculated \cite{giamarchi,
  fisher} that a new phase of matter exists under this condition: The
Bose glass (BG).  A series of experiments in 1D \cite{fallani,
  deissler} showed that such a phase exists also for a quasiperiodic
potential. This experimental research has been accompanied by 
a corpus of analytical and numerical articles \cite{roux,
  prokofev, anna, modugno, fontanesi1, fontanesi2, cetoli} that helped
explaining the properties of the glassy phase. In particular, it is
predicted that the BG should appear in more than one dimension, and
experiments are being done to verify this.

In this Letter we want to investigate the BG phase in
a 2D QC lattice. The work of Guidoni \emph{et al}
\cite{guidoni97, guidoni99} opened up for the possibility of creating an
optical QC using an appropriate configuration of lasers. In
particular, a configuration with five beams creates a ten-fold symmetric
structure, similar to the Penrose tiling \cite{penrose,gardner}. 
For a noninteracting system, this configuration is known to have eigenstates
which are neither extended (like Bloch functions), nor exponentially
localized. Surprisingly, few works have been done to study the
interacting system. Among these, a numerical study by Sanchez-Palencia
and Santos \cite{santos} has shown a that a Penrose-like potential is
able to inhibit the diffusion of a BEC. We wish to fill the gap in the
literature and show that the quasiperiodic lattice is compatible with
a gapless insulating phase, as in the Bose glass.
\begin{figure}[tbp]
\centering
\includegraphics[width=3.1in,clip]{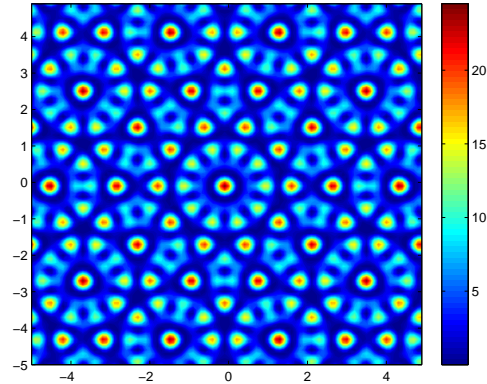}
\caption{Penrose potential generated by 5 opposing laser beams, as in
  Eq. (\ref{penrose}) with $V_0=1\,E_R$. In our configuration, the
  polarization of the laser beams is directed towards the reader.
  }
\label{penrose}
\end{figure}

The optical realization of a QC gives a choice of the sign of the
potential: the lattice can be a quasiperiodic array of wells or
peaks. We choose the latter, which can be realized with a blue detuned lattice, because in such a
potential there are no potential wells and a Mott insulating phase can not exist. 
The possible loss of coherence of the bosons will then depend solely on 
localization effects due to the quasiperiodicity.
\begin{figure*}[ht!]
\centering
\includegraphics[width=0.98\textwidth,clip]{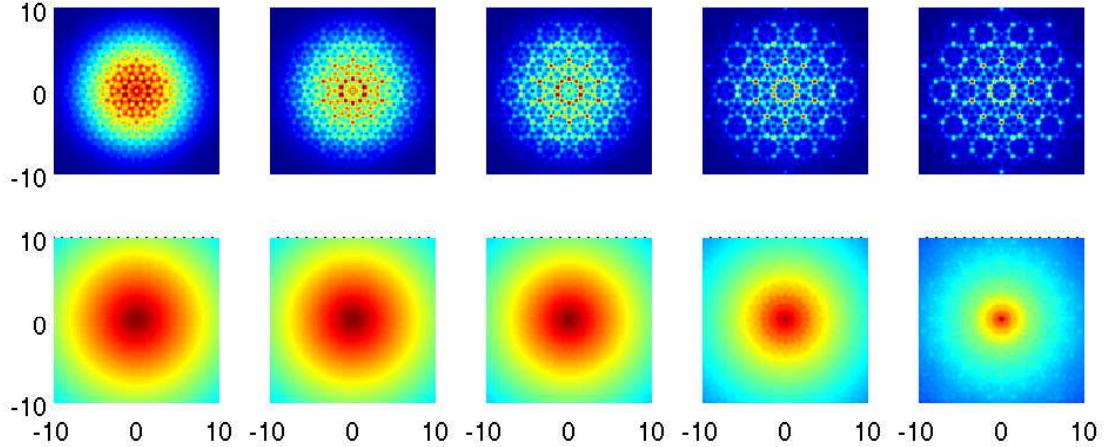}
\caption{Condensate density (top row) and spatial average of the
  correlation function as in Eq (\ref{spatial_aver}), for $g=10$ with
  $V_0=1,2,3,4$ and $4.5\,E_\mathrm{R}$. In this colorscale blue is
  the lowest value, red is the highest. }
\label{corr}
\end{figure*}

\section{Penrose tiling}
A quasiperiodic optical lattice can be generated in 1D by means of
superimposing two laser beams with incommensurate wavelengths
\cite{fallani}. In two dimensions it is possible to create a tenfold symmetric 
quasicrystal structure using five laser beams of equal intensity,
lying in the same plane with an angle $2\,\pi/5$ between them. The
resulting potential is given by the expression \cite{grimm}
\begin{eqnarray}
V_\mathrm{QP}(x,y) = V_0\,\left|\sum_i \mathcal{E}_i\,\mathbf{\epsilon}_i
                               e^{-i\,(\mathbf{k_i}\cdot\mathbf{r}+ \phi_i)}\right|^2
\,,
\end{eqnarray}
where $\mathbf{r} = (x,y)$, $\mathcal{E}_i$ is the relative
dimensionless intensity of each laser beam, $k_i$, $\phi_i$, and
$\mathbf{\epsilon}_i$ are the corresponding wavevector, phase, and
polarization. $V_0$ is an overall intensity. The wavelength $d=
2\,\pi/|k|$ of the lattice sets the characteristic length of the
system. In the following, we measure all the energies in terms of the
recoil energy $E_R= \hbar/m\,d^2$, and the unit length is $d$.
Moreover, we choose $\mathcal{E}_i=1$, $\phi_i=0$, and the various
polarization vectors point all in the same direction, perpendicular to
the $xy$ plane. In Fig. \ref{penrose} we plot the resulting potential
for $V_0=1\,E_R$. This potential appears as a quasiperiodic structure
of peaks in the plane.

We consider a quasiperiodic potential combined with a harmonic
confinement, so that the external potential in the $xy$ plane is
\begin{eqnarray}
V(x,y)= V_\mathrm{QC}(x,y) + \frac12 \omega^2 (x^2+y^2)
\,.
\end{eqnarray}

\section{Bogoliubov approach}

The 2D gas of bosons considered here will in the absence of a
quasiperiodic potential form a Bose-Einstein condensate; in the
anticipated Bose-glass phase, long-range coherence will be lost so
that the gas can be considered a quasicondensate \cite{popov}. In
both cases, the system is accurately described by a (quasi-)
condensate wavefunction $\psi_0(x,y)$, accompanied by quadratic
fluctuations described by Bogoliubov modes \cite{mora,alkhawaja,wu}.
The (quasi-) condensate wavefunction is governed by the well-known 2D
Gross-Pitaevskii equation (GP2D)
\begin{equation}
\frac12 \nabla^2 \psi_0 + g\,|\psi_0|^2\,\psi_0 + V\,\psi_0 = \mu\,\psi_0
\,.
\end{equation}
In this 2D approximation, the wavefunction is supposed to
be constant in the direction perpendicular to the $xy$ plane,
extending for a length equal to $a_\perp$. The value of $g=
g^{(\mathrm{3D})}/a_\perp$ is the scaled interaction strength among
the particles in 2D, $V(x,y)$ is the external potential, and $\mu$ is
the chemical potential. The density of the (quasi-) condensate is
given by $n=|\psi_0|^2$. Up to second order in the many-body
Hamiltonian the atoms outside the (quasi-) condensate occupy excited
states which are determined by solving the Bogoliubov equations
\begin{eqnarray} 
\left( -\frac12 \nabla^2 + v_j + 2\,n\,g- \mu \right) \, u_j
    - n\,g\,v_j &=& \omega_j\,u_j
\nonumber\\
\left( -\frac12 \nabla^2 + v_j + 2\,n\,g- \mu \right) \, v_j
    - n\,g\,u_j &=& -\omega_j\,v_j
\,.
\end{eqnarray}
where, as usual, the normalization $\int u_i\,u_j- v_i\,v_j=
\delta_{ij}$ is enforced. As was found in Ref. \cite{mora, cetoli}, the
one-body correlation function can be expressed in terms of the
Bogoliubov excitations as
\begin{eqnarray}
\ln\,g_1(\mathbf{r},\mathbf{r}')
&=&
\ln \langle \hat{\psi}^\dagger(\mathbf{r})\, \hat{\psi}(\mathbf{r}')\rangle - \ln \sqrt{n\,n'}
\nonumber \\
&=&
-\frac{1}{2}
\,
\sum_{j\ne0} \Big\{ |\frac{v_j}{\sqrt{n}}-\frac{v_j'}{\sqrt{n'}}|^2 
\nonumber\\
&+& \; N_j\, \left[ |\frac{u_j}{\sqrt{n}}-\frac{u_j'}{\sqrt{n'}}|^2 + |\frac{v_j}{\sqrt{n}}-\frac{v_j'}{\sqrt{n'}}|^2 \right]  \Big\}\,,
\end{eqnarray}
provided $\psi_0$ is real. In this formula, $\hat{\psi}$ is the many-body 
boson operator, and for brevity we write $u_j,v_j,n$ and
$u'_j,v'_j,n'$ instead of
$u_j(\mathbf{r}),v_j(\mathbf{r}),n(\mathbf{r})$ and
$u'_j(\mathbf{r}),v'_j(\mathbf{r}),n'(\mathbf{r})$. $N_j$ is the
occupation number of the $j^{th}$ excited state, as determined by the
Bose distribution. In current experiments, this occupation can be made so
small that it can be neglected with respect to the depletion from the
(quasi-) condensate given by the interaction. In the following we work
with $N_j=0$ (as for vanishing temperature) and the only contribution
to the correlation is the first term in the sum.

We remark that the Bogoliubov approach is a gapless {\it ansatz} for
the excitations in a Bose system. This approach works for the gapless Bose
glass phase, but it fails to detect the Mott insulator transition
because the latter phase has a gap. The choice of our
potential inhibits the formation of a Mott phase, and the only
insulating phase in the system is provided by the quasiperiodic
pattern of the lattice.

\begin{figure}[tbp]
\includegraphics[height=2.5in,clip]{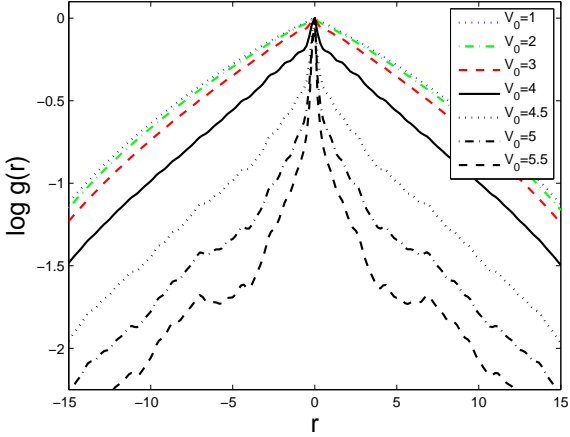}
\caption{Logarithm of the rotation average of the correlation
  function, as in Eq. (\ref{eq:rot}).}
\label{rot}
\end{figure}

We find the ground state of the GP2D by using an imaginary time
evolution with the Fourier split operator method. Subsequently, we
diagonalize the Bogoliubov equations by employing the ARPACK libraries
with Tchebychev polynomial acceleration \cite{cheby}. We choose a
system size of $L=80\,d$, and the square grid has $480$ points per
side. The matrix to diagonalize has therefore the dimensions of
$2\times 480^2\sim 4.6\times 10^5$ points per side. 
This matrix is not banded, since we have chosen to represent the
Laplacian using the Fourier transform. Therefore, in order to have
convergence in reasonable time, even with the polynomial
acceleration, we must make a cutoff at $100$ eigenvectors. We shall
see that this cutoff is adequate.

We consider a system with $g=10$; this relatively weak interaction can be 
realized with, e.g., a  gas 
of $\sim10^5$ atoms, confined in a transverse dimension $a_\perp\sim
100 \mathrm{nm}$, with a scattering length of $\sim
10^{-4}\mathrm{nm}$, which we propose could be realized using Feshbach 
resonances. 
The harmonic potential strength in the
$xy$ plane is $\omega=8\times 10^{-2}\,E_R$. 
The reason for these parameter values is numerical limitations; in 
principle, there is nothing to prevent the BG transition to occur also 
for larger values of $g$. 
Also, the numerical results are limited to 
$V_0 \leq 5.75\,E_\mathrm{R}$: Beyond this parameter regime, an unattainably high 
spatial resolution was required in order to correctly represent the 
mode functions. 

The top row of Fig.
\ref{corr} plots the density of the condensate while changing the
strength of the lattice from $V_0=1\,E_\mathrm{R}$ to
$4.5\,E_\mathrm{R}$. In the bottom row we plot instead the spatial
average of the correlation function \cite{naraschewsky}
\begin{eqnarray} \label{spatial_aver}
g(\mathbf{r}) 
= \frac1V \int d\mathbf{R}\,\langle \hat{\psi}^\dagger(\mathbf{R}+\mathbf{r})\,
                                    \hat{\psi}(\mathbf{R})\rangle\,.
\end{eqnarray}
This quantity is directly related to what is imaged in experiments;
the density after time of flight is to a good approximation equal to the
momentum space density $\rho_k$,
\begin{eqnarray}
\rho_k 
 &=& \frac1V \, \int d\mathbf{r}\,d\mathbf{R}
                     \,e^{i\,\mathbf{k}\cdot\mathbf{r}}
                     \,g(\mathbf{r})
\,.
\end{eqnarray}
As we can see, for low values of $V_0$ the characteristic width
of $g(\mathbf{r})$ is the entire condensate, indistinguishable
from the average correlation function of a pure condensate in a
harmonic potential. When $V_0\sim 3\,E_R$ this correlation function
starts to shrink, and eventually decays rapidly when getting farther
from the center.

This behavior is particularly evident when we plot the angular
average of $g(\mathbf{r})$
\begin{equation}\label{eq:rot}
g(r) = \int \frac{d\theta}{2\,\pi} \,g(\mathbf{r},\theta)
\,,
\end{equation}
as is shown in Fig. \ref{rot}. When a weak lattice is present, the
logarithm of the correlation function decays on a scale comparable
with the total size of the system. Upon increasing the lattice, a
central peak appears where the logarithm of the correlation decays
linearly, on a scale shorter than the system size.  This is one
of the signatures of a crossover to an insulating phase, as reported
by Deissler \emph{et al} \cite{deissler}.

\section{Elementary excitations}
It is known that the Bose glass phase is caused by low-lying
excitations that can flip the phase of the quasicondensate with a
little amount of energy. One of the signatures of the glassy phase is
the progressive lowering of the excitation spectrum. 

\begin{figure}[tbp]
\centering
\includegraphics[width=3.3in,clip]{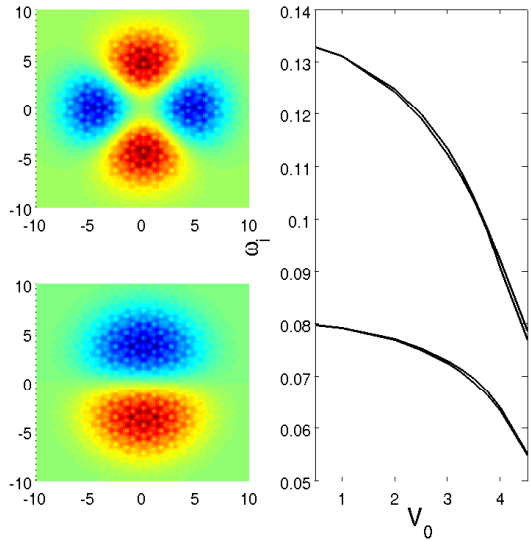}
\caption{Dipolar and quadrupolar excitations. Right panel: 
The four lowest excitation energies, in units of $E_R=\hbar^2/m\,d^2$, as 
functions of the lattice strength $V_0$. 
Lower left: Bogoliubov amplitude $v_1$ for the first excited mode; 
upper left: Bogoliubov amplitude $v_3$ for the third excited mode, 
both computed for the case 
$V_0=1\,E_\mathrm{R}.$
}
\label{exc}
\end{figure}

In Fig. \ref{exc} we plot the energies of the lowest
dipole and quadrupole excitations. We see that the raising of the
quasiperiodic lattice separates each of the modes into two branches. More
importantly, the energy of the excitations drops significantly. 
These modes can be excited by imposing dipolar and quadrupolar
deformations in the confining potential. In particular, a strain can
be given to the quasicondensate by a slight change in the harmonic
confinement,
\begin{equation}
\frac12 \omega^2\,(x^2+y^2)
\rightarrow
 \frac12 (\omega+\epsilon)^2\,x^2
+
 \frac12 (\omega-\epsilon)^2\,y^2
\,.
\end{equation}
Such a deformation will predominantly set off an excitation with an irrotational velocity field. 
For $V_0\sim 0$, the
characteristic frequency of this mode is seen to be $\tilde \omega\approx
0.13\,E_R \approx 1.6\omega$. This is the scissor mode, described in Ref. \cite{odelin}. 
Using a hydrodynamical approach, Ref.
\cite{odelin} shows that a superfluid admits quadrupolar excitations
(scissor modes) in a harmonic potential, and the lowest frequency for
this type of excitation is in the strongly interacting limit equal to 
$\sqrt{2}\,\omega$; for our finite value of $g$, it is slightly higher 
at $V_0=0$.  
Another important result of Ref.
\cite{odelin} is that a normal fluid would dampen this excitation in a
finite time. We argue that the disappearance of the scissor mode
provides a way to determine the phase transition to the Bose glass,
since when the system is completely normal the quadrupolar excitations
are short-lived.

In order to compute the lifetime of the scissor modes 
one should compute the imaginary part of the
stress-tensor--stress-tensor response function. This cannot be done in
the simple Bogoliubov approximation, because even for a uniform system
the response function involves an integral that is ultraviolet
divergent. In order to lay down a microscopic theory of the
quadrupolar excitations a renormalization approach seems compulsory,
but such a work is outside the scope of the current letter.

\section{Superfluid fraction}
\begin{figure}[tbp]
\centering
\includegraphics[width=3.35in,clip]{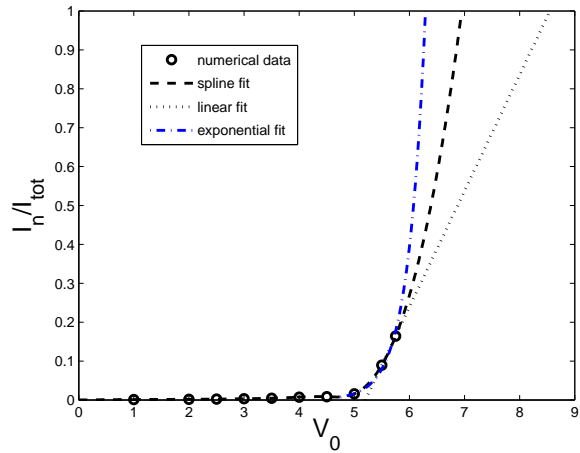}
\caption{Normal fraction of the system as function of $V_0$ (in units
  of $E_R=\hbar^2/m\,d^2$), according to Eq. (\ref{eq:normal}). The
  circles correspond to the ratio $I/I_\mathrm{tot}$ computed
  according to Eq. (\ref{eq:normal}). In order to estimate where the
  superfluid part disappears, we plot a spline interpolation of the
  whole data, a linear fit of the last two points, and an exponential
  fit of the last three points.}
\label{normal}
\end{figure}
The clearest signature of the Bose glass is the fact that the
superfluid fraction of the density vanishes, and the system acts as a
normal fluid. The superfluid fraction is the subset of the fluid that
can move only according an irrotational velocity field, while the rest
of the gas constitutes the normal fraction. Inspired by the work in
Ref. \cite{pines}, we compute the normal fraction by looking at the
response to a rotation of the system. The external potential can be
rotated at a small angular velocity $\Omega$. In the rotating frame
the Hamiltonian of the system appears as 
\begin{equation}
\hat{H}'= \hat{H} - \Omega\,\hat{L}_z 
\,,
\end{equation}
where the angular momentum operator is (using hats on second-quantized operators but not on first-quantized operators)
\begin{equation}
\hat{L}_z(\mathbf{r},t)
 = L_z(\mathbf{r})\,\left.\hat\psi^\dagger(\mathbf{r}',t')\,\hat\psi(\mathbf{r},t)
                               \right|_{\mathbf{r}'\rightarrow \mathbf{r},
                                        t'\rightarrow t}
\,,
\end{equation}
and $\mathbf{r}=(x,y)$, so that $L_z(\mathbf{r})= -i\,\hbar\,(x\,\partial_{y}-
y\,\partial_{x})$. 

If the stirring is slow enough, superfluid vortices cannot be
generated, and only the normal part takes part in the rotation of the
fluid. In the limit of vanishing $\Omega$, the increase in energy due
to the rotation is given by
\begin{equation}
E_{\Omega} - E_0 = \frac12 \, I_n \,\Omega^2
\,,
\end{equation}
where $I_n$ is the moment of inertia of the normal part. Using
standard perturbation theory, the normal moment of inertia is given by
second order perturbation theory. Expressed as integrals over the
imaginary time $\tau=-i\,t$, it reads
\begin{eqnarray}
I_n
 = \frac2{\beta}\int d\tau_1 \,d\tau_2\,
   \int d\mathbf{r}_1\, d\mathbf{r}_2\, 
    \langle \hat{L}_z(\mathbf{r}_1,\tau_1) \hat{L}_z(\mathbf{r}_2,\tau_2) \rangle 
\,,
\end{eqnarray}
with $\beta=1/T$, the inverse of the temperature. Notice that this
expression is the $q=0,\omega=0$ limit of the
angular-momentum--angular-momentum response function, as described by
Pines and Nozi{\`e}res \cite{pines}. By applying Wick's theorem we
obtain, in the Bogoliubov approximation
\begin{eqnarray} \label{eq:normal}
I_n
 &=& 2\,\sum_{i \ne j} \frac{\mathcal{I}_{ij}}{\omega_i+\omega_j} 
     +
     \sum_{i} \frac{\mathcal{I}_{ii}}{\omega_i}
\,,
\end{eqnarray}
with
\begin{eqnarray}
\mathcal{I}_{ij}
&=&  
 (\int v_i\,L_z\,u_j)\,(\int u_j\,L_z\,v_i)
\nonumber\\
 &+&
 (\int v_i\,L_z\,u_j)\,(\int u_i\,L_z\,v_j)
\,,
\end{eqnarray}
where the last integrals are only over the spatial coordinates, since
the (imaginary) time has been integrated out in the evaluation of the
Matsubara frequencies \cite{cetoli}.  The normal fraction is given by
the ratio $I_n/I_\mathrm{tot}$, where $I_\mathrm{tot}$ is the total
momentum of inertia. This ratio is plotted in Fig. \ref{normal}.
Note that the sum over the excitations is limited by the cutoff in
the diagonalization procedure. On the other hand, we checked that this
restriction of the number of modes does not alter the result: 
The highest 10\%
of the states included were seen to contribute less than $1\%$ of
the sum. This is expected, since the Bose glass phase depends on the
lowest lying excited modes.

According to our analysis, the normal part starts to increase when the
correlation function shows an exponential decay on a scale of the
whole system ($V_0\sim 5 E_\mathrm{R}$). We cannot see the full
transition because no numerical results could not be obtained for $V_0
> 5.75\,E_\mathrm{R}$, as discussed above.  However, an assortment of
extrapolation methods -- Fig.\ \ref{normal} shows results of spline,
exponential, and linear extrapolation -- indicate that the superfluid
part should vanish between $V_0 =6\,E_\mathrm{R}$ and $V_0
=9\,E_\mathrm{R}$. Moreover, note that the transition to the glassy
phase is not sharp, but it appears as a smooth crossover from
superfluid to insulator.  This effect is due to the finite dimensions
of the system, and for the 1D Bose glass it has been experimentally
observed in Ref. \cite{deissler}.

\section{Conclusions}
We have seen evidence suggesting that a 
a blue detuned optical quasicrystal generates a
phase transition in a Bose gas, from a superfluid to an insulating
phase. The signatures of this transition are an exponential decay of
the correlation function and an increase in the normal part of the
gas. Since we have chosen a gapless {\it ansatz} for the excitations,
and a the blue detuned lattice without potential wells, 
we know that this phase is not a Mott insulator. The transition is due
to the specific shape of the potential and the localization effects it
causes on the Bose gas. Summing up, the quasiperiodic pattern of the
lattice leads to a normal, gapless state, compatible with the
description of the Bose glass.

\begin{acknowledgments}
  We thank Alice Bezett, Ben Deissler, Claude Dion, and Harri
  M\"akel\"a for insightful discussions. 
This project was financially supported by the Swedish Research Council, 
Vetenskapsr{\aa}det, and was conducted using the resources of High 
Performance Computing Center North (HPC2N).
\end{acknowledgments}


\end{document}